**Stratégies de Gestion de la Chaleur et Performances Sportives de Haut Niveau :**

**Eclairage Psycho-physiologique et Recommandations Appliquées**

Heat Management Strategies and High Level Sports Performance: Psycho-Physiological

Insights and Applied Recommendations

Stratégies de gestion de la chaleur et recommandations appliquées

N. Robin[1], E. Hermand[2], V. Hatchi[1] & O. Hue[1].

[1]Laboratoire "Adaptation au Climat Tropical, Exercice & Santé", Faculté des Sciences du Sport de Pointe-à-Pitre, Université des Antilles, Campus Fouillole, BP 592, 97159, Pointe à Pitre Cedex, France.

[2]ULR7369 - URePSSS - Unité de Recherche Pluridisciplinaire Sport Santé Société, Univ. Littoral Côte d'Opale, Univ. Artois, Univ. Lille, F- 59279 Dunkerque, France.

Nicolas Robin, Tel +(0033)5 90 48 31 73 ; Fax +(0033)5 90 48 31 79 ; robin.nicolas@hotmail.fr

Résumé

(1) Objectifs : Cet article apporte un éclairage sur les différentes stratégies de gestion de la chaleur, usuelles ou innovantes, afin d'analyser celles qui seraient les plus adaptées et les plus efficaces chez des athlètes devant participer à des compétitions dans des environnements humides et/ou chauds.

(2) Actualités : Les Jeux Olympiques de Paris en 2024 se dérouleront du 26 juillet au 8 septembre avec un risque élevé pour les athlètes de pratiquer leur sport à des températures élevées, imposant de fait des contraintes physiologiques (cardio-vasculaires, ventilatoires, thermorégulatoires…) et psychologiques (fatigue mentale précoce, baisse de motivation, inconfort…) pouvant ainsi largement impacter négativement leur performance.

(3) Perspectives et projets : Pour « performer » en environnement chaud, il est aujourd'hui recommandé d'avoir recours à des stratégies, notamment une acclimatation active qui favorise des adaptations physiologiques mais aussi psychologiques. De même, les techniques de gestion de fluides et de refroidissement « physiques » ont des effets potentiellement bénéfiques sur des facteurs physiologiques mais leurs conséquences psychologiques sont encore peu connues et doivent être investiguées. Enfin, des stratégies mentales (fixation d'objectifs, imagerie mentale, dialogue interne positif, musique…) ou entrainements cognitifs en environnement chaud peuvent limiter les contre-performances dans ces conditions. Les effets de la combinaison de techniques physiques et mentales, ainsi que des stratégies innovantes comme la suggestion au froid sont également en cours d'investigation.

(4) Conclusion : Pour chacune des stratégies présentées, les travaux scientifiques ont permis l'élaboration de recommandations pratiques à l'intention des athlètes, des entraîneurs et des préparateurs mentaux, afin de leur permettre d'anticiper physiologiquement et psychologiquement les effets d'une hygrométrie et ou d'une température élevée.

*Mots clés* : Stress thermique, sport, refroidissement, innovation, athlète,

Heat Management Strategies and High Level Sports Performance: Psycho-Physiological Insights and Applied Recommendations

Abstract

(1) Objectives : This article sheds light on the different heat stress management strategies, common or innovative, in order to analyze those that would be the most suitable and effective for athletes who have to compete in humid and/or hot environments.

(2) News: The Paris summer Olympics Games in 2024 will take place from July 26$^{th}$ to September 8$^{th}$ with a high risk for athletes to practice their sport at high temperatures, thereby imposing physiological (cardiovascular, ventilatory, thermoregulatory…) and psychological (early mental fatigue, decreased motivation, discomfort…) which can have a major negative impact on their performance.

(3) Prospects and projects : To perform in a hot environment, it is now recommended to use strategies, in particular active acclimatization wich promotes physiological but also psychological adaptations. Similarly, fluid management and cooling techniques have potentially beneficial effects on physiological factors but their psychological consequences are still poorly understood and need to be investigated. Finally, mental strategies (goal setting, mental imagery, positive self-talk, music…) or cognitive training in the heat can limit poor performance in this condition. The effects of combining physical and mental techniques, as well as innovative strategies such as cold suggestion are also being investigated.

(4) Conclusion: For each strategies presented, the scientific work has enabled the development of practical recommendations for athletes, coaches and mental trainers in order to allow them to physiologically and psychologically anticipate the effects of high relative humidity and/or high temperature.

Stratégies de Gestion de la Chaleur et Performances Sportives de Haut Niveau : Eclairage Psycho-physiologique et Recommandations Appliquées

Des compétitions sportives internationales ont régulièrement lieu en condition de températures humides et/ou élevées (e.g., Jeux Olympiques de Rio 2016, Championnat du monde d'athlétisme de Doha 2019) exposant les athlètes à des environnements pouvant induire des contraintes physiologiques susceptibles de limiter notamment les performances aérobies [1, 2]. Mais elles peuvent aussi engendrer des conséquences psychologiques susceptibles d'augmenter l'inconfort thermique, d'impacter négativement les affects ainsi que le fonctionnement cognitif, ce qui peut nuire aux performances sportives de très haut niveau dans de nombreuses disciplines sportives (e.g., sports d'équipe, sports duels, sports de précision…) [3-5]. Par exemple, nous pouvons relever que les Jeux Olympiques (JO) et Para-Olympique de Paris 2024 seront réalisés à une période de l'année (i.e., saison d'été d'un climat tempéré selon la classification de Köppen) pendant laquelle les températures atteignent fréquemment 33°C et peuvent potentiellement dépasser 42°C à certains horaires, comme lors d'un épisode de canicule en juillet 2019 (Bulletin climatique de météo France, 2019). C'est pourquoi il est probable que les JO de 2024 ou certaines de ses épreuves puissent se dérouler en condition de Température Elevée (TE). Il est donc recommandé de planifier des préparations spécifiques en amont, mais aussi des stratégies physiologiques (e.g., acclimatation, refroidissement) et psychologiques (gestion des émotions, imagerie mentale, suggestion) pendant, voire entre les épreuves ou rencontres sportives, notamment lorsqu'elles s'enchaînent à des intervalles variés, afin que les athlètes optimisent leurs chances de « performer » [6].

Cette synthèse a pour but d'évoquer les stratégies de gestion d'une température environnementale élevée, des plus usuelles aux plus innovantes, afin d'analyser celles qui pourraient être adaptées et efficaces et donner des recommandations appliquées utiles pour les athlètes, coaches et préparateurs mentaux dans le cadre de la préparation à des événements sportifs futurs tels que les JO de Paris en 2024. En effet, réaliser des performances de haut niveau nécessite une combinaison de facteurs physiologiques, techniques, tactiques mais également psychologiques [7] pour espérer atteindre le maximum de ses capacités du moment. Alors que la plupart des travaux de recherche ont porté sur les conséquences physiologiques d'une TE (e.g., hyperthermie, dépassement des capacités thermorégulatoires), peu d'études se sont intéressées aux conséquences psychologiques (affects, motivation, inconfort, cognition) d'un environnement chaud [4, 8-9] alors que celles-ci peuvent fortement impacter les performances sportives des athlètes de haut niveau. Nous évoquerons donc, dans une première partie, les principales conséquences d'une TE. De même, alors que des stratégies de lutte contre la chaleur au moyen de techniques de gestion des fluides, d'acclimatation ou de refroidissement « physiques » ont été préalablement testées et ont montré des effets potentiellement bénéfiques sur des facteurs physiologiques [10-11], les conséquences psychologiques de ces stratégies sont peu connues. Ces dernières doivent impérativement être prises en compte afin d'éviter toute contre-performance en compétition et ainsi permettre à l'athlète d'être en mesure d'atteindre un niveau de performance optimal [7]. Après avoir abordé les principales stratégies usuellement utilisées en contexte sportif, nous évoquerons également des travaux de recherches innovants portant sur les effets de techniques de « refroidissement » mentales (i.e., suggestion au froid, imagerie mentale), de gestion des conséquences d'une TE (e.g., pleine conscience, dialogue interne), ou d'inhibition de celles-ci (e.g., musique). De même, nous proposerons des pistes de recherches suggérant le recours à des entraînements cognitifs spécifiques, notamment sur les processus attentionnels ou les temps de réaction, en amont des compétitions. Ceux-ci permettraient, en anticipant les conséquences négatives d'une TE, de garder le plus longtemps possible un haut niveau de fonctionnement cognitif (e.g., concentration, attention sélective, attention soutenue, inhibition) malgré les contraintes thermiques. Enfin, comme évoqué dans de précédents travaux de recherches (e.g., [1, 2]), nous suggèrerons aux athlètes et entraîneurs de recourir à des phases de préparation en condition de TE mais aussi dans des environnements encore plus « extrêmes » combinant notamment une chaleur élevée avec un haut degré d'humidité comme le climat tropical [1]. L'idée est de bénéficier à la fois d'adaptations physiologiques (e.g., [13]) et psychologiques (e.g., [7]) permettant de mieux « s'adapter » aux conséquences d'une TE en compétition et favoriser ainsi des performances sportives de haut niveau.

**Conséquences d'une Température Elevée**

Des facteurs exogènes comme la chaleur peuvent influencer négativement l'activité physiologique en imposant des contraintes aux systèmes cardiovasculaire et ventilatoire, se traduisant notamment par une augmentation des fréquences cardiaque et respiratoire, liées au maintien de l'homéostasie [6, 14-15]. La TE engendre ainsi des difficultés de dissipation de chaleur corporelle pouvant augmenter la température centrale, occasionner des dysfonctionnements physiologiques [2] ainsi qu'un épuisement cognitif précoce [16, 17]. De même, [18] ont montré que l'augmentation de la ventilation en condition de TE pouvait réduire le flux sanguin cérébral ce qui pouvait impacter négativement les processus d'élimination de la chaleur au niveau du cerveau et engendrer une contrainte cognitive supplémentaire pouvant nuire au traitement de certaines tâches cognitives complexes [19, 20] et favoriser l'apparition de fatigue mentale [21]. De plus, et comparativement à un environnement dans lequel la température est neutre, une TE peut altérer l'état émotionnel [22], diminuer le confort thermique [11, 23] ou augmenter les affects négatifs tels que la colère ou l'agressivité [24].

Nous intéressant plus particulièrement aux performances sportives, un exercice à des températures ambiantes élevées fait subir aux corps des athlètes des contraintes physiologiques plus grandes qu'un exercice du même type dans des conditions tempérées, pouvant entrainer une diminution des performances aérobies [25]. En effet, à des intensités élevées d'exercice en condition de TE, on observe une augmentation de la température corporelle directement liée à des altérations cardiovasculaires, neuromusculaires, métaboliques, immunologiques pouvant entraîner une fatigue prématurée (e.g., [13-14, 26-27]). Les effets d'une TE sur les performances anaérobie (e.g., sprint) sont plus contrastés, allant d'une légère amélioration sur des sprints isolés à une nette détérioration sur des sprints répétés [28, 29]. Alors que la plupart des études ont porté sur l'influence de la TE sur les facteurs physiologiques et notamment pour les activités de type aérobie [1, 2] et anaérobie [29], d'autres travaux de recherche ont porté sur l'influence de l'exercice sur des facteurs psychologiques comme la cognition [3]. Ces derniers ont souligné que de nombreux exercices prolongés (e.g., sports duels, sports collectifs, courses de longue durée) nécessitent le maintien d'un fonctionnement cognitif élevé, comme la prise de décision, pour « performer » [30]. En effet, dans les situations où des actions sont dirigées vers un but à atteindre, les résultats reposent généralement sur la capacité des athlètes à réguler de manière précise et répétée leurs comportements en fonction de l'objectif visé. Par exemple, les sportifs peuvent être amenés à interpréter des informations visuelles, auditives, ou proprioceptives spécifiques à leur environnement, à initier des ajustements comportementaux immédiats en réponse à un ou plusieurs adversaires ou partenaires, à localiser et tenter d'atteindre des cibles stables ou en mouvement, à faire évoluer leurs stratégies en fonction du contexte ou à tenir compte des décisions des arbitres. Un tel ensemble de circonstances montre la nécessité d'un fonctionnement cognitif optimal au cours de toute activité sportive mais à laquelle l'ajout d'une TE est susceptible d'engendrer des contraintes (e.g., baisse des ressources attentionnelles disponibles) et donc d'affecter négativement les performances [5, 21]. La TE a aussi pour conséquence de favoriser l'apparition d'une fatigue précoce [31], d'amplifier l'effort perçu [32], d'augmenter l'inconfort thermique [33], de diminuer la durée des exercices [34], mais aussi d'allonger la durée de réalisation d'une course et d'augmenter le nombre d'abandons dans des épreuves de longues durées [35]. Selon [6] différentes stratégies préventives de lutte contre les effets négatifs de la TE ont été étudiées comme l'acclimatation ou l'ingestion de fluide pour limiter la déshydratation [36-37] afin notamment d'atténuer le risque d'hyperthermie. Nous allons à présent évoquer ces différentes stratégies usuellement utilisées en condition de TE.

**Stratégies Usuelles de Gestion d'une Température Elevée**

### Acclimatation en Environnement Chaud

De nombreuses études ont tenté de vérifier s'il y avait un intérêt à recourir à des stratégies d'acclimatation, passive ou active, dans le cadre de la pratique sportive [13]. Il a ainsi été montré que l'acclimatation à une TE entraînait des adaptations physiologiques telles qu'une réduction de la consommation d'oxygène à une puissance donnée, une économie du glycogène musculaire, une réduction de lactate sanguin à une puissance donnée, une augmentation de la force produite, une augmentation de la sudation, une diminution de la concentration en sel de cette dernière, une augmentation du volume plasmatique et une amélioration de l'efficacité du myocarde [12, 38-39], tout en limitant les risques d'abandons [40] ainsi que les problèmes de santé liés à la chaleur [2]. Ces derniers ont ajouté que la répétition d'exercices en condition de TE favorisait l'acclimatation active des athlètes à la chaleur, via une série d'adaptations physiologiques (e.g., diminution de la fréquence cardiaque) améliorant la thermorégulation et réduisant les contraintes physiologiques (i.e., diminution de la température centrale) pour des exercices réalisés à une intensité donnée. De même, [38] ont montré que les adaptations physiologiques, obtenues après une période d'acclimatation à une TE, avaient permis d'améliorer à postériori les performances aérobies aussi bien en climat tempéré (13°C) qu'en condition de chaleur (38°C). Les auteurs ont proposé que des protocoles d'acclimatation à la chaleur pourraient être utilisés pour compléter les routines d'entraînement des athlètes. Selon [13], la plupart des adaptations physiologiques sont obtenues au bout de quelques jours en condition de TE. Cependant, la capacité à réaliser un exercice physique en condition de chaleur semble s'améliorer de façon optimale après une à deux semaines dans les conditions en question avec une pratique physique régulière, ce qui amène des auteurs (e.g., [13, 40]) à suggérer d'être sur les lieux de compétitions au moins 14 jours avant le début des épreuves (voir Figure 1).

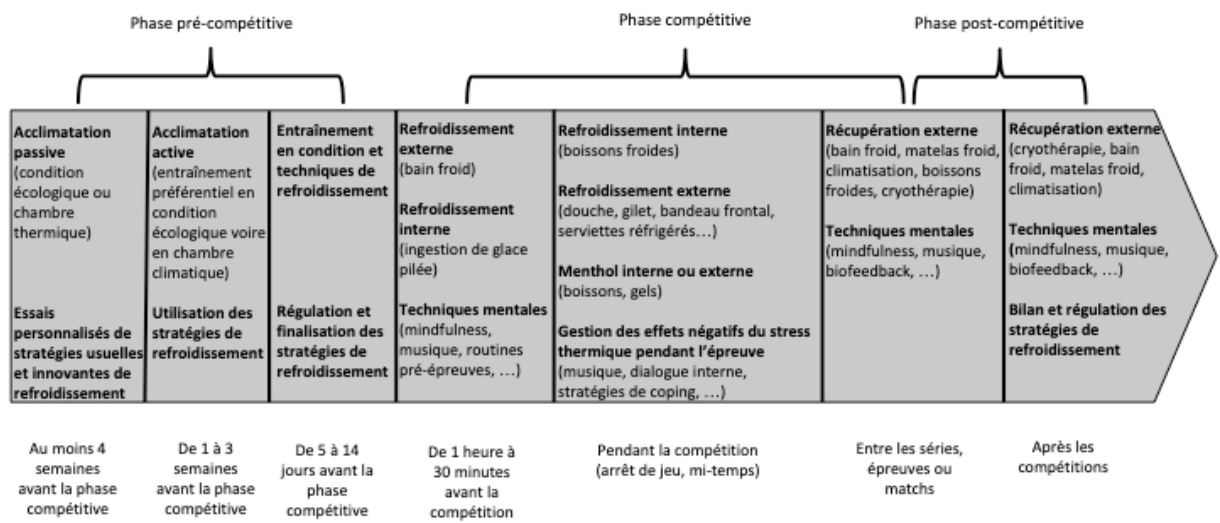

*Figure 1.* Stratégies d'adaptation à la chaleur et de récupération en compétitions sportives

Il est aussi important de noter que l'acclimatation ne porte pas seulement sur les facteurs physiologiques mais peut aussi influencer les facteurs psychologiques. L'hyperthermie causée conjointement par la TE et l'exercice physique augmente l'inconfort thermique ainsi que les affects négatifs et réduit les affects positifs [41] ce qui peut affecter la capacité d'exercice [42] ainsi que les performances cognitives [24]. Néanmoins, il a été mis en évidence que l'acclimatation à la chaleur pouvait améliorer la sensation et le confort thermiques, chez les athlètes de sports collectifs et d'endurance, mesurés au cours d'entraînements et de compétitions réalisés en condition de TE [6]. De plus, l'acclimatation peut aussi limiter les effets négatifs de l'hyperthermie sur les performances dans des tâches cognitives d'attention [43], de planification [41] ou de type psychomotrice [44], ce qui devrait favoriser les performances des athlètes dans les activités sportives sollicitant ces processus. Enfin, [7] ont suggéré que des athlètes habitués à supporter des contraintes (e.g., intensité et durée d'exercice élevées) en condition de TE percevraient la difficulté de l'exercice physique comme moins importante, en comparaison avec des athlètes moins habitués à de telles conditions ce qui permettrait notamment de réduire les abandons dans les épreuves de longues durées (e.g., marche, marathon, triathlon). C'est la raison pour laquelle une acclimatation en climat tropical combinant une TE et un haut degré d'hygrométrie (plus de 70% d'humidité relative) permettrait à la fois de bénéficier des adaptations physiologiques [45] (mais favoriserait aussi une adaptation psychologique à ce climat « extrême », permettant de mieux supporter (e.g., moins d'inconfort, affects et facteurs motivationnels moins négativement impactés) un environnement chaud et sec [1]. A ce titre, quelques travaux de recherche ont mis en évidence une tolérance au travail supérieure chez des soldats entraînés [46] après une acclimatation en climat tropical ou une baisse de l'effort perçu et de l'inconfort thermique [47] suite à un entraînement en environnement chaud et humide dans lequel l'état d'hydratation des militaires était contrôlé. Cependant, il est très important de souligner, qu'en plus de surveiller l'état d'hydratation des athlètes, il sera important de veiller à ce que ces derniers ne basculent pas dans le surentrainement par un excès de charge, dont celle environnementale, trop important et n'aient pas d'excès de confiance (en condition extrême) car il est potentiellement dangereux de chercher à repousser ses limites.

**Gestion de l'Etat d'Hydratation**

En condition de TE, on observe une augmentation de la sudation (transpiration) causée par des processus physiologiques de thermorégulation [27]. Celle-ci induit ainsi une déshydratation corporelle, s'accentuant au décours de l'exercice physique [48]. La régulation de la réponse sudorale à l'exercice est un des deux facteurs principaux du maintien de l'équilibre des fluides, l'autre étant l'ingestion de liquide [6]. Il est important de noter que la sensation de soif peut ne pas être un bon indicateur du statut d'hydratation de l'athlète et que l'ingestion de liquide basée sur l'envie de boire peut malgré tout conduire à une déshydratation pendant l'exercice [49]. Les travaux de recherches portant sur les conséquences de la déshydratation sur la performance sportive ont pointé des résultats contradictoires. Alors que des auteurs suggèrent de ne pas dépasser 2% de perte de masse corporelle liée à un déficit en eau [50], d'autres auteurs ont mis en évidence qu'une déshydratation aiguë allant jusqu'à 3% de perte de masse corporelle, avec ou sans perception de la soif, n'avait aucune influence sur la performance dans un contre-la-montre de 20 km en cyclisme en condition de TE [51]. Ainsi, certains athlètes, notamment dans les courses pédestres de longue durée, favoriseraient la déshydratation au bénéfice d'une masse corporelle moindre pour gagner en vélocité [35]. Cependant, le lien entre déshydratation et performance sportive nécessite d'être approfondi par d'autres travaux en condition de terrain car les conséquences d'une déshydratation trop importante en TE peuvent provoquer des crampes [52], mais également des malaises pouvant entraîner une hyperthermie maligne d'exercice (e.g., coup de chaleur d'exercice) et causer la mort [53]. C'est pourquoi, nous invitons les athlètes à la plus grande prudence et leur conseillons de tester et d'intégrer la gestion de liquide (i.e., contenu, volume, fréquence, température) dans leurs routines d'entraînement et de performance.

En ce qui concerne le fonctionnement cognitif, des travaux de recherche ont montré que la déshydratation impactait négativement les performances dans des tâches impliquant les fonctions exécutives, les temps de réaction et les ressources attentionnelles [54-56] et notamment entraînait des difficultés de concentration, augmentait la fatigue et réduisait la volonté de poursuivre un effort physique [57-58]. De plus, [3] ont indiqué qu'outre son effet néfaste direct sur les performances cognitives, la déshydratation pourrait également moduler indirectement la cognition par son impact sur les mécanismes de thermorégulation et sur l'hyperthermie. C'est pourquoi, [59] recommande d'anticiper les pertes hydriques et suggère de commencer l'ingestion de liquide avant d'atteindre 1 à 2% de la perte de masse corporelle, c'est-à-dire avant que la déshydratation induite par le TE et/ou l'exercice ne commence à altérer de manière marquée les processus cognitifs. Par contre, il est important de noter qu'il faut éviter l'hyperhydratation qui n'améliore ni la thermorégulation ni les performances [35]. De plus, [3] suggèrent de compléter les boissons ingérées par les athlètes avec des électrolytes mais aussi du sodium qui favorisera l'absorption des liquides par l'organisme et des glucides qui permettront de répondre à la charge cognitive supplémentaire en condition de TE [60]. L'ingestion de liquide peut se faire au moyen de boissons fraîches ou de glace pilée. Ces dernières pourraient également être considérées comme des techniques de refroidissement, mais elles peuvent causer des nausées, de l'inconfort, des troubles gastriques [61], ou être peu efficaces [62]. C'est pourquoi, des auteurs (e.g., [63-64]) suggèrent de les utiliser en complément d'autres techniques de refroidissement (e.g., gilet de froid) dont les effets sur la baisse de température centrale seront amplifiés lorsque l'athlète sera dans un état euhydraté par rapport à un état déshydraté [65]. Nous allons à présent aborder les différentes stratégies de refroidissement.

**Stratégies de Refroidissement Physique**

Des stratégies de refroidissement visant à atténuer les effets négatifs de la chaleur et réduisant à la fois l'hyperthermie ainsi que les baisses de performances cognitives peuvent être particulièrement utiles pour préserver la performance et naturellement la santé des athlètes [3, 37]. Selon [66], chez un individu non acclimaté l'augmentation de la température centrale induite par la TE et/ou l'exercice est associée à des contraintes physiologiques plus importantes (e.g., hypovolémie, augmentation du taux de sudation, fréquences cardiaque et respiratoire plus élevées). Quelles que soient les stratégies de refroidissement et leur moment d'utilisation (i.e., avant, pendant et/ou après un exercice physique), l'objectif est de ralentir l'accumulation de chaleur endogène et de limiter ainsi l'augmentation de la température centrale des athlètes. Des auteurs comme [64, 67] ont proposé de classer les techniques de refroidissement en deux catégories : les stratégies internes et les stratégies externes. Parmi les stratégies internes, on retrouve l'ingestion d'eau froide (entre 0 et 10°C) et de glace pilée, combinées ou non avec du menthol. Le menthol permet notamment, via la stimulation des thermorécepteurs au froid (i.e., TRPM8), de déclencher une sensation de froid et de modifier les perceptions thermiques [68-70]. De plus, des travaux de recherche ont montré que le menthol permettait une meilleure performance en condition de TE [61, 71-74). Le menthol doit cependant être prudemment et précisément dosé (de 0,01% à 0,1%) dans les boissons car l'information thermique transmise par les récepteurs au froid ainsi stimulés est susceptible de biaiser l'analyse des situations par le système nerveux central [69, 75]. Néanmoins, un nombre grandissant d'études souligne le maintien de la température corporelle interne en TE concomitant à de meilleures performances aérobies, malgré une production de chaleur corporelle mécanique (contractions musculaires) théoriquement plus importante [27, 61, 71-72], lors d'une utilisation interne (ingestion) ou externe (application cutanée) de menthol. Selon [76], les stratégies de refroidissement internes, par ingestion de liquide ou de glace pilée, ont l'avantage de pouvoir être utilisées en compétition : elles limitent l'augmentation des températures centrales et corticales et la baisse des performances aérobies en condition de TE [73, 77], tout en hydratant les athlètes. Cependant, il est important de noter que mal utilisées, les stratégies de refroidissement internes peuvent être contreproductives et générer de l'inconfort, des troubles gastriques, des allergies ou des céphalées [78].

Les stratégies de refroidissement externes (e.g., bains, douches, vêtements réfrigérés ou pack de glace) sont généralement utilisées avant les compétitions comme stratégies de pré-refroidissement (ou pre-cooling) permettant de lutter contre la TE [14, 79] mais plus rarement pendant les épreuves en per-refroidissement (ou per-cooling) comme dans l'étude de [16]. Ce sont des stratégies efficaces qui permettent de diminuer la température centrale avant l'exercice et d'augmenter la capacité de stockage de la chaleur corporelle. De plus, en réduisant la température cutanée, le refroidissement externe favorise la réduction du flux sanguin cutané et les tensions cardiaques et cérébrales associées à la perfusion de la peau. En diminuant la température centrale et/ou la température cutanée, ces stratégies de refroidissement peuvent réduire les informations afférentes issues des thermo, chimio et barorécepteurs endogènes et diminuer les réponses thermorégulatrices comme la transpiration, permettant de freiner la déshydratation [3] et préservant le niveau de performance dans de nombreuses activités sportives [80]. Il est important de noter qu'au cours d'un exercice physique, les stratégies externes sont plus difficilement utilisables, soit à cause de durées limitées d'utilisations possibles (e.g., serviettes humides ou poches de glace lors des temps-morts, des changements de côtés ou d'adversaires), soit à cause des règlements des compétitions qui interdisent l'usage de certaines techniques comme les gilets, cagoules réfrigérantes et limitent ainsi les possibilités d'obtenir un refroidissement externe efficace [40, 81] ou uniquement sur des surfaces limitées (e.g., bandeau frontal réfrigéré, bandanas) comme évoqué par [7]. Enfin, les stratégies de refroidissement comme l'immersion en eau froide ou les douches froides peuvent être utilisées après les efforts (i.e., post-refroidissement ou post-cooling), notamment lors des périodes de récupération pour améliorer les performances répétées d'endurance en condition de TE [82-84]. De même, des travaux de recherche en cours s'intéressent aux potentiels bénéfices des matelas réfrigérés : ils permettraient d'augmenter la durée et qualité du sommeil et ainsi favoriser la récupération des athlètes.

    Alors qu'indépendamment du type d'intervention, un effet ergogène du pré-refroidissement sur les performances d'endurance en condition de TE est généralement rapporté [37, 80], les preuves de l'efficacité des stratégies de refroidissement sur les performances cognitives restent limitées. Cependant, ces stratégies pourraient être appropriées

en fonction du moment de leur mise en œuvre [3]. Selon ces derniers, les bénéfices physiologiques apportés par les stratégies de refroidissement pourraient présenter un intérêt au regard de la théorie de l'espace global de travail [85] selon laquelle on dispose de ressources cognitives limitées [9]. En effet, [86] ont montré l'utilisation de davantage de ressources cognitives en condition de TE, pour rester concentré, dans des tâches d'attention. Il a ainsi été envisagé que la réduction de la contrainte thermique augmenterait la disponibilité des ressources attentionnelles qui pourraient ensuite être utilisées pour faire face aux demandes cognitives liées à l'exercice et imposées par le stress thermique [4, 87]. De même, alors qu'une TE induit généralement une augmentation de la température cutanée qui peut causer de l'inconfort, du déplaisir et détériorer l'activité corticale [24], la limitation de la hausse de cette température au moyen de gilet réfrigéré, de poches de glace ou de serviettes froides pourrait permettre le refroidissement de la peau et ainsi favoriser les performances des athlètes dans des tâches exigeantes sur le plan cognitif [14], lorsque la réglementation le permet. Enfin, il semble que des combinaisons de techniques de refroidissement internes (e.g., ingestion de boisson fraîche) et externes (e.g., gilet, poche de glace), utilisées en pré- et per-refroidissement pourraient protéger temporairement les athlètes réalisant des performances cognitivo-motrices prolongées en condition de TE [64]. En revanche, de récents travaux de recherche suggèrent de tester individuellement, et avant les compétitions, la ou les stratégies de refroidissement (internes et externes) ainsi que leurs moments d'utilisation (pré-, per- voire post-refroidissement). En effet, il sera important de veiller à ce que les techniques utilisées aient des effets positifs sur des facteurs physiologiques (e.g., températures centrales et/ou cutanées) sans occasionner d'effets négatifs sur des facteurs psychologiques (e.g., inconfort et sensation thermique augmentée, inconfort des équipements de refroidissement, augmentation des affects négatifs, déconcentration) afin de favoriser la performance des athlètes en condition de TE [7]. Selon ces derniers, il est aussi possible d'avoir recours à d'autres types de stratégies de refroidissement que nous allons à présent évoquer.

**Stratégies Mentales et Cognitives en Contexte de Température Elevée**

**Stratégies Mentales Usuelles**

En condition de chaleur, les facteurs psychologiques comme la motivation, la volonté de poursuivre un effort et les perceptions thermiques peuvent être négativement impactés avant tout changement physiologique mesurable [88] ce qui peut impacter les performances cognitives et physiques. C'est pourquoi, [89] a proposé qu'une performance sportive, en condition de TE, pourrait être améliorée au moyen de l'entraînement d'une ou de plusieurs habiletés psychologiques en amont de la pratique physique. Par exemple, [90] ont testé si l'entraînement d'un ensemble de quatre habiletés psychologiques (i.e., fixation d'objectifs, régulation de l'excitation, imagerie mentale et dialogue interne positif) pouvait augmenter la distance parcourue au cours de trois « contre la montre » de 90 minutes réalisés en condition de TE (i.e., 30°C, 40%rH). Les auteurs ont montré que cet entraînement psychologique avait permis d'augmenter la distance parcourue de 8% ce qui représentait une amélioration de 1.15 kilomètre, notamment en limitant la tentation de réduire l'intensité de l'exercice pendant ces courses d'effort maximal. Les résultats de cette étude montrant l'intérêt d'un entraînement psychologique en condition de TE sont certes intéressants mais ils ne permettent pas de déterminer si l'entraînement d'une seule de ces habiletés n'aurait pas permis d'obtenir des résultats similaires. C'est la raison pour laquelle, [89] ont supposé qu'une seule de ces stratégies, à savoir le dialogue interne positif, pourrait être bénéfique à la performance en condition de TE, sachant qu'il avait déjà été montré un effet positif de cette habileté mentale, en condition thermique neutre, sur la performance dans des exercices d'endurance [91-92] ainsi que sur des processus cognitifs comme l'attention, la concentration et l'humeur [93-94]. Ainsi, [89] ont ainsi fait l'hypothèse que deux semaines d'entraînement au dialogue interne positif aurait des effets bénéfiques et améliorerait aussi bien les capacités d'endurances (i.e., épreuve de temps maximum de pédalage à une intensité donnée) que le fonctionnement cognitif (i.e., fonctions exécutives, temps de réaction, mémoire de travail) en condition de TE de part une régulation top-down de la performance. Les résultats de cette étude montrent que le dialogue interne positif a permis une amélioration significative de la durée maximale de pédalage ainsi qu'une amélioration de la vitesse et de la précision dans les tâches cognitives impliquant les fonctions exécutives (e.g., mémoire spatiale).

D'autres études ont montré, en condition neutre, que des stimulations externes, comme la vidéo, le soutien social ou encore l'écoute de musique, facilitaient la réduction ou la suppression de l'effort perçu et de la douleur physique [95-97]. En ce qui concerne le cas de l'écoute de musique, celle-ci peut améliorer l'exécution lors de l'effort, en réduisant la sensation de fatigue, en augmentant les niveaux d'excitation, en encourageant la coordination et/ou la synchronisation motrice et en augmentant la relaxation [98-99]. Pendant l'exercice en musique, la réduction de la sensation de fatigue peut être liée à une attention sélective résultant de la capacité limitée de traitement des informations [100]. Ainsi, l'écoute de la musique empêche le sportif d'être attentif simultanément aux sensations de fatigue dues à l'effort. Ce modèle est connu sous le nom de modèle de traitement parallèle. Il a été proposé que ce mécanisme n'ait de l'influence qu'à faible intensité d'exercice lorsque les signaux externes peuvent concurrencer les signaux internes [101]. Ces effets peuvent être dus à la distraction [102-105]. Ainsi, la musique présente un stimulus qui permet de détourner (temporairement) l'attention sur des stimuli externes (e.g., [99]) et, par la même occasion, de moins porter son attention sur les sensations internes (douleur, fatigue, inconfort thermique). Or, la chaleur affecte les sensations internes en provoquant un inconfort thermique qui amenuise la performance, par notamment un effort perçu plus élevé qu'en condition neutre, tout au long de l'effort (e.g., [106]). Le taux d'accumulation de chaleur permet une réduction anticipée de l'intensité de l'exercice (par rapport à un milieu neutre) pendant la course à un niveau fixe d'effort perçu [107]. Ainsi, la musique crescendo entraînant une augmentation progressive du stimulus musique tout au long de l'effort (i.e., une liste musicale dans laquelle le tempo augmente au fur-et-à-mesure) devrait être d'autant plus efficace en condition de TE pour diminuer les sensations désagréables et augmenter la performance due à la déviation de l'attention. Comme précédemment évoqué, le danger de ce type d'intervention pourrait être le dépassement des seuils physiologiques, l'hyperthermie et les non prises en compte des signaux d'alertes pouvant amener l'athlète à dépasser ses limites et se mettre en danger. C'est pourquoi pour des efforts de longue durée en environnement humide et/ou chaud, la musique synchrone (i.e., musique dans laquelle le tempo est synchronisé avec le mouvement) pourrait s'avérer plus adaptée lors de l'effort afin d'éviter que le sportif ne se retrouve en surrégime.

Très peu de travaux de recherche ont étudié l'effet de la musique sur un effort réalisé en condition de TE. Par exemple, [108] ont examiné si l'écoute de musique synchrone affectait le maintien de la performance de course, la lactatémie sanguine, la fréquence cardiaque, le confort thermique et le niveau d'effort perçu des athlètes dans des conditions chaudes et humides. Les auteurs ont sollicité 12 coureurs masculins qui ont effectué deux essais dans des conditions contrôlées (31°C et 70% d'humidité) avec ou sans musique. Cette étude a montré que le temps de course jusqu'à épuisement en condition de musique synchrone était 66,59% plus long (moyenne = 376.5 sec.) que le temps de course en absence de musique (moyenne = 226.0 sec.). De plus, les scores de RPE étaient significativement plus faibles pour la condition musicale synchrone (vs. condition non musicale) à chaque instant (15, 30, 45 et 60 min.) sauf après épuisement où le score était identique. Les auteurs suggèrent que l'écoute de la musique synchrone est bénéfique pour la performance ainsi que pour l'effort perçu et ceci est d'autant plus le cas lorsque les athlètes se trouvent dans des environnements chauds et humides. L'usage de la musique comme technique de motivation et de régulation de l'intensité peut donc être une voie prometteuse et innovante pour toutes les disciplines aérobies que ce soit à l'entraînement, lors des phases d'acclimatation voire même au cours des compétitions (sous réserves de compatibilité avec les règlements).

**Stratégies Innovantes**

De récents travaux de recherche ont proposé que des stratégies mentales de refroidissement pourraient être envisagées en condition de TE [7]. Ces stratégies devraient permettre de limiter les effets négatifs de la chaleur sur les facteurs psychologiques tels que la motivation à poursuivre un effort. C'est pourquoi le recours à des techniques mentales aidant les athlètes à gérer « psychologiquement » la TE semble prometteur. En effet, [109] ont testé si une intervention axée sur la suggestion au froid, en condition de TE, avait des effets spécifiques sur des marqueurs psychologiques (confort thermique, sensations thermiques, affects) en comparaison avec une absence d'intervention. Les auteurs ont observé que l'écoute d'une bande sonore de dix minutes suggérant une baignade dans une eau très froide, comme celle d'une rivière, permettait d'améliorer le confort thermique, de trouver la condition de TE moins chaude et d'augmenter la motivation des participants en comparaison avec une absence d'intervention. Cependant, bien que prometteuse, l'efficacité de cette stratégie était limitée à une dizaine de minutes et n'était pas réalisée en condition de pratique physique, ce qui a amené les auteurs à suggérer d'avoir recours à d'autres techniques comme la méditation de pleine conscience qui pourrait être adaptée à des conditions de TE même si à notre connaissance aucune étude n'a montré son efficacité en condition de stress thermique. La méditation de pleine conscience comporte trois étapes : 1) la lucidité avec prise de conscience des pensées, des émotions et des sensations corporelles du moment présent ;
2) l'acceptation de par une attitude non critique vis-à-vis des pensées, des émotions et des sensations corporelles actuelles, et 3) la re-concentration sur les objectifs, décisions à prendre ou sur l'effort [110]. Bien qu'elle ne se soit pas déroulée dans des conditions environnementales particulières, l'étude de [111] a montré qu'une intervention de pleine conscience de sept semaines modifiait la façon dont les athlètes de haut niveau traitaient les informations intéroceptives et augmentait leur capacité à réguler l'anxiété liée aux sensations désagréables. Des expérimentations seront prochainement réalisées dans notre laboratoire pour tester l'effet potentiellement bénéfique de cette stratégie sur les performances sportives réalisées en condition de TE.

**Stratégies d'Entraînement Cognitif**

Tout comme pour les entraînements physiques, il semble que des entraînements cognitifs spécifiques puissent être bénéfiques aux athlètes [11 2] notamment lorsqu'il s'agira d'optimiser ses performances sportives en condition de chaleur. Selon [31], de nombreux sports de longue durée nécessitent le maintien d'un fonctionnement cognitif optimal (e.g., prise de décision, attention soutenue, concentration) que ce soit pour l'atteinte de performance que pour limiter les risques de blessures [113]. En effet, dans les sports dans lesquels les actions sont dirigées vers un objectif, la performance des athlètes repose sur leur capacité à réguler de manière répétée et précise leur comportement en fonction de cet objectif [3]. Ainsi, afin optimiser leur préparation et retarder la fatigue mentale qui peut apparaître suite à des périodes prolongées d'activités cognitives exigeantes [114] et qui peut être exacerbée par l'exercice physique [31] et/ou la TE [21], nous suggèrerons aux athlètes de recourir à des entraînements des fonctions exécutives, attentionnelles ou perceptivo-motrices dans des environnements chauds. En effet, [44] ont observé une amélioration des performances psychomotrices chez des pilotes de rallye qui s'étaient entraînés avec des simulateurs de courses pendant quatre jours en condition de TE. De même, il existe des dispositifs d'entraînements cognitifs comme les cellules lumineuses dont l'allumage est géré par des programmes informatiques installés sur des tablettes. Ce dispositif, qui permet d'améliorer le temps de réaction de choix et la précision motrice, est fréquemment utilisé par les entraîneurs, préparateurs physiques et athlètes de sports duels ou collectifs, en condition neutre et il semblerait pertinent de réaliser également des séquences d'entraînements dans des environnements chauds afin d'aider à retarder les effets négatifs d'une TE. De plus, il existe des logiciels qui consistent à entraîner l'attention et la vigilance des athlètes en leur demandant de suivre des cibles, en trois dimensions (i.e., 3D) en mouvement, dans des environnements virtuels statiques ou dynamiques et qui permettent de développer la prise de décision, de mieux gérer les flux optiques et d'augmenter la durée de concentration. Le fait d'avoir recours à des entraînements cognitifs de ce type, en condition TE et avant les compétitions sportives, devrait permettre aux athlètes de retarder l'apparition de la fatigue mentale et d'augmenter leur durée de concentration et de vigilance. Ces entraînements devraient ainsi permettre d'augmenter la durée de fonctionnement cognitif optimal des athlètes et donc favoriseraient la performance lors des épreuves mobilisant des ressources cognitivo-motrices (e.g., sports duels, collectifs, de précision) réalisées en condition de chaleur.

**Conclusion**

Lorsque l'on souhaite « performer » en condition de température élevée, il est fortement recommandé de recourir à des stratégies d'acclimatation et également de refroidissement physique (interne ou externe) qui peuvent être administrées avant, pendant, voir entre et après les épreuves sportives. Ces stratégies peuvent être utilisées seules mais nous recommandons de les combiner, ou en cas d'impossibilité de règlement d'avoir recours à des stratégies psychologiques de refroidissement comme la suggestion au froid ou des techniques mentales de lutte contre les conséquences d'une TE comme la méditation de pleine conscience, le dialogue interne positif ou une combinaison de ces dernières. Ces stratégies sont à considérer comme des complément potentiellement intéressants mais qui doivent être envisagées avec précaution étant donné les capacités de dépassement des athlètes de haut niveau et donc d'augmentation des risques de contre performance. De même, étant donné que les bénéfices d'une ou plusieurs stratégies de lutte contre la chaleur, ou la préférence d'utilisation de celle(s)-ci, soit spécifique à chaque individu, nous encourageons les délégations à tester en milieu écologique (i.e., acclimatation naturelle) et de façon active (i.e., avec la réalisation d'exercices physiques) quelle(s) stratégie(s) serai(en)t la(es) plus efficace(s) pour chaque athlète. Les effets bénéfiques doivent à la fois concerner les adaptations physiologiques que les perceptions psychologiques. Nous recommandons également que ces stratégies soient intégrées très tôt dans les routines des athlètes afin de permettre à ces derniers de se familiariser avec leur utilisation et augmenter leur efficacité. Enfin, nous suggérons aux athlètes de haut niveau et aux coaches de recourir à des entraînements cognitifs spécifiques afin de lutter plus efficacement contre les effets négatifs d'une TE sur les processus cognitifs des athlètes et ainsi favoriser l'atteinte de performances sportives optimales lors des compétitions réalisées en condition de chaleur.

## Références